\documentclass[12pt]{article}
\begin{document}

\newcommand{\la} {\langle}
\newcommand{\ra} {\rangle}

\title{Does Kolmogorov mean field theory become exact for turbulence above some
critical dimension?}

\author{Mark Nelkin}

\maketitle

Physics Department, New York University, New York, NY 10003,\\
and Levich Institute, CCNY, New York, NY 10031 USA\\
Electronic address: Mark.Nelkin@nyu.edu

\begin{abstract}
I give three different arguments for an upper critical dimension $d_{max}>3$ above which the 1941 Kolmogorov mean field theory becomes essentially exact, and anomalous scaling vanishes.  The first argument concerns the number of degrees of freedom in a turbulent flow and indicates that $d_{max}=4$.  The second argument is a naive estimate of dangerous fluctuations, and also suggests that $d_{max}=4$.  The third argument is related to a known critical point of the GOY shell model when the amplitude of back energy transfer becomes small.  This third argument does not give a numerical value for $d_{max}$.  None of these arguments bears any known relationship to any of the others nor to the generally accepted qualitative physical picture of the dynamical origin of anomalous scaling in turbulence.  Despite this, the three arguments together suggest that the suggestion of an upper critical dimension should be taken seriously.
\end{abstract}

\pagebreak

The 1941 Kolmogorov theory (K41)  \cite{frisch} is a good first approximation to the universal statistics of the small scale velocity fluctuations in high Reynolds number turbulence.  In this theory, the probability distribution function (pdf) of the velocity difference $\Delta u(r)$ between two points separated by a distance $r$ is self-similar and nearly Gaussian.  It is well established from experiment and direct numerical simulation (DNS) that this self-similarity is only approximate in three dimensional turbulence. As the scale separation $r$ is decreased, the tails of the pdf become enhanced.  The deviations from self-similarity are known as anomalous scaling.

Most studies of anomalous scaling in the inertial range have concentrated on the longitudinal structure functions,

\begin{equation}
D_p(r)=\la [u(x+r)-u(x)]^p \ra = \la [\Delta u(r)]^p \ra,
\end{equation}
which are the moments of the velocity difference $\Delta u(r)$ between two points separated by a distance $r$.  The term ``longitudinal" refers to the component of the velocity difference along the line joining the two points.  These structure functions are characterized by power-law behavior in the inertial range with anomalous scaling exponents $\zeta_p$ defined by
\begin{equation}
D_p(r)=Const. \la \epsilon \ra^{p/3} \; r^{p/3} (r/L)^{(\zeta_p-p/3)}.
\label{zeta}
\end{equation}

The ``normal" scaling of K41 corresponds to $\zeta_p=p/3$.  The observed values of $\zeta_p$ for $p>4$ are smaller than the K41 values corresponding to an intermittent form for the pdf where large values of $\Delta u(r)$ are increasingly enhanced as $r$ is decreased.  This phenomenon is often called the internal intermittency of turbulent flows.

It was natural to suggest in the 1970's \cite{nelkin74} that K41 theory played the role of a mean field theory for turbulence.  By analogy with equilibrium critical phenomena, it might be that this mean field theory could become exact either below some critical spatial dimension $d_{min}$ or above some critical dimension $d_{max}$.  There have been many speculations about this possibility, but until recently these speculations have not led to any serious physical insight.  I suggest here that recent developments might change this situation.

A very important development has been the observation that the inverse cascade of energy in two dimensional turbulence shows normal K41 scaling.  This has been observed in the laboratory by Paret and Tabeling \cite{paret98}. It was first seen in DNS by Smith and Yakhot \cite{smith93}, explored in more detail by Boffetta et al. \cite{boffetta}, and has been given some theoretical understanding by Yakhot \cite{yakhot2d}.  Recently Yakhot \cite{yakhotnew} has revived the idea of continuously varying the dimension $d$ between $d=3$ and $d=2$, and has suggested that there is a lower critical dimension $d_{min}$  at which the direction of the energy flux changes sign.  This idea has been given important and interesting support within the GOY shell model in a recent paper by Giuliani, Jensen and Yakhot \cite{giuliani}.  They find that intermittency does not vanish near this critical dimension even though it is absent in $d=2$. I return to a more quantitative discussion of this model later in the present paper.

It is generally agreed, as first suggested by Kraichnan \cite{rhk74}, that anomalous scaling arises from a competition between a spatially local cascade, and a spatially non-local dynamical averaging of coupled local cascades.  This spatial mixing is intimately associated with the non-local pressure term in the Navier-Stokes equation.  This competition is expected to be a smooth function of spatial dimension with intermittency decreasing as the spatial dimension increases.  There are indications, however, that a crossover dimension $d_{max}$ may exist with K41 mean field theory becoming exact for $d>d_{max}$.  One such indication comes from a calculation of the number of degrees of freedom in a turbulent flow.  Using the $\beta$ model of fractally homogeneous turbulence \cite{frisch}, Kraichnan \cite{rhk85} showed that anomalous scaling decreased the number of degrees of freedom for $d<4$, but increased the number of degrees of freedom for $d>4$.  Meneveau and Nelkin \cite{meneveau89} extended this result to a general multifractal model.  The crossover occurs because the effects of anomalous scaling on the density of modes in the inertial range competes  with the effects of anomalous scaling on the fluctuating viscous cutoff, and this competition depends on the spatial dimension $d$.

It is appealing to think that there is some physics in the above estimate.  Anomalous scaling is often thought to be associated with coherent structures in the flow, and therefore with a reduction of randomness, and a reduction in the number of degrees of freedom.  At first sight it does not seem physically plausible for anomalous scaling to increase the number of degrees of freedom.  Furthermore it is important to remember that anomalous scaling is a property of a statistically stationary state, and need not have any direct dynamical origin.  On the other hand, there is no known reason why the number of degrees of freedom in a turbulent flow should be of direct physical interest.  Furthermore the calculation of the crossover dimension depends on the mechanism of viscous dissipation.  If the viscous term proportional to $\nu k^2$ in the Navier-Stokes equation is replaced by a hypo-viscous term proportional to $\lambda k^p$ with $p < 2$, Nelkin and Meneveau \cite{nelkin99} showed that the same crossover effect calculated for ordinary viscosity at $d=4$ occurs for $d=3$ when $p=5/3$.  This suggests looking at a DNS of three dimensional turbulence with varying hypoviscous index $p$ to see if any reduction of intermittency can be seen when $p < 2$.

There is another  argument which suggests a crossover at $d=4$.  The anomalous scaling in turbulence is associated with anomalous fluctuations of the energy dissipation rate which are, in turn, related to the fluctuations of the nonlinear energy transfer through Kolmogorov's refined similarity hypothesis \cite{k62}.  To identify ``dangerous" fluctuations, we need a locally defined dynamical variable which is a reasonable surrogate for the nonlinear energy transfer.  A possible choice is

\begin{equation}
T({\bf r})= {\eta}^2 (\partial u/\partial x)^3
\end{equation}
where $\eta$ is the dissipation length scale, and $u$ is the local velocity.  This variable is ``regularized" in the sense that its average value is constant in the limit of zero viscosity.  Since the scaling dimension of the velocity derivative is $(2/3)$, the scaling dimension of $T({\bf r})$ is $6(2/3)=4$.  This suggests an upper critical dimension $d_c=4$.
The preceding argument was given in an earlier paper  \cite{nelkin78}.  It is not clear if it is valid, but it is not obviously wrong. There is a quite different sense in which this argument might be ``dangerous."  It admittedly arises from a search for a dynamical variable which might show the desired crossover properties.

Finally there are interesting crossover properties observed in the GOY \cite{gledser} \cite{ohkitani} shell model of turbulence.  Since this model is a one dimensional dynamical model, it does not make any explicit reference to the spatial dimension $d$, but there are reasonable arguments \cite{giuliani} which suggest how the parameters in the model might vary with $d$.  The starting point is a set of wave numbers $k_n=k_0 2^n$ and an associated set of complex amplitudes $u_n$ of the velocity field.  Each amplitude interacts with nearest and next-nearest neighbor shells, and the corresponding set of coupled ODE's takes the form:

\begin{equation}
du_n/dt = -\nu k_n^2 u_n + f_n + C_n
\end{equation}
where the nonlinear coupling term is given by

\begin{equation}
C_n=i k_n [a_n u^*_{n+1} u^*_{n+2} + (b_n/2)  u^*_{n-1} u^*_{n+1} + (c_n/4) u^*_{n-1} u^*_{n-2}].
\end{equation}
I take my notation from  \cite{giuliani}. 

The values of the coupling constants are fixed by imposing conserved quantities.  To conserve the total energy when $f_n=\nu=0$, we require $a_n+b_{n+1}+c_{n+2}=0$.  The time scale is fixed by the condition that $a_n=1$ leaving the free parameter $\delta$ by defining the coupling constants as

\begin{equation}
a_n=1, b_n=-\delta, c_n=-(1-\delta).
\end{equation}

It has been known for some time that the choice $\delta=1/2$ gives good agreement between the temporal intermittency in the GOY model and the observed spatial intermittency of thee dimensional turbulence.  It was pointed out by Kadanoff et al \cite{kadanoff95} that this choice of $\delta$ gives a second conserved quadratic quantity which corresponds loosely to helicity.  It is not at all understood, however, why the temporal intermittency in this one dimensional model should be related to the spatial intermittency in real three dimensional turbulent flows.  

The GOY model can be generalized to allow the scale ratio $\lambda$ between adjacent shells to be different than $2$.  It was pointed out by Kandanoff et al \cite{kadanoff95} that the agreement with experiment remains good for varying values of $\lambda$ as long as the second conserved quadratic quantity continues to correspond to helicity.

This observation strongly suggests that there could be physical interest in varying the parameter $\delta$.  The recent paper by Giuliani, Jensen and Yakhot \cite{giuliani} shows that as $\delta$ is increased, there is a critical value at $\delta=1$ at which the direction of the energy cascade changes. The value $\delta=5/4$ corresponds to enstrophy as the second conserved quadratic quantity, and might be related to the inverse cascade in two dimensional turbulence. This supports the idea that increasing $\delta$ from $\delta=1/2$ corresponds in some sense to reducing the dimension in turbulence from $d=3$ towards $d=2$.

In this paper I would like to emphasize what happens when $\delta$ is reduced below $\delta=1/2$. In the same spirit as above, this might correspond to increasing the spatial dimension to $d>3$.  It was shown by Biferale et al \cite{biferale95} that there is a critical value of $\delta\approx 0.38$. For values of $\delta$ smaller than this, the K41 fixed point becomes stable and intermittency vanishes.  Further studies have shown that the properties of this critical point are quite subtle and can be affected by viscosity \cite{schorgorfer}, but I choose here to ignore these subtleties.  Clearly this suggests that there is an upper critical dimension $d_{max}>3$ above which intermittency vanishes, and the K41 mean field theory becomes essentially correct.

I should emphasize that the dynamical mechanism in the GOY model for the vanishing of intermittency is quite different than the conventionally assumed competition between local cascade and spatial mixing \cite{rhk74}.  The one dimensional GOY model does not consider space at all.  As $\delta$ is decreased, the relative amplitude of backwards energy transfer is decreased, and this reaches a critical value below which intermittency is no longer maintained.  

An interesting physical argument concerning the origin of this critical point has been given by Biferale and Kerr \cite{bifkerr}.  They suggest that the fluctuations in the second conserved quadratic quantity become ``dangerous" when $\delta$ is larger than this critical value.  The fluctuations are ``dangerous" when they are dominated by small scale fluctuations in the K41 solution.  This estimate of the transition value of $\delta$ agrees well with numerical experiment for varying values of the scale ratio $\lambda$.

In conclusion, I have suggested that there is a critical dimension $d_{max}>3$ above which anomalous scaling in turbulence vanishes and the K41 mean field theory becomes essentially exact.  Counting degrees of freedom, and naive arguments about ``dangerous" fluctuations suggest that $d_{max}=4$.  The existence of a transition to a stable K41 fixed point in the one dimensional GOY shell model has been known for some time, and it suggests that $d_{max}>3$, but gives no indication of its numerical value.  In all of these arguments, there is no connection made to the usual picture of the dynamical origin of intermittency nor is there any known connection among the three arguments presented here. Despite these limitations, it is suggestive that there is important physics in turbulence for $d>3$ which deserves serious further study.

\end{document}